\documentclass{svproc}
\usepackage{url}

\usepackage{graphicx}
\usepackage{amsmath}
\usepackage{dirtytalk}

\newcommand{\papertitle}{Quantifying life quality as walkability on urban networks: the case of Budapest}
\newcommand{\papercitation}{This is a pre-print of \say{\papertitle} published In: Cherifi H., Gaito S., Mendes J., Moro E., Rocha L. (eds) \emph{Complex Networks and Their Applications VIII.} COMPLEX NETWORKS 2019. Studies in Computational Intelligence, vol 882. The final version is available at: \url{https://doi.org/10.1007/978-3-030-36683-4_72}} 

\begin{document}

\mainmatter              
\title{\papertitle}

\titlerunning{Life quality on urban networks}  
%
\author{Luis Guillermo Natera Orozco \inst{1} \and D\'{a}vid Deritei\inst{1}
Anna Vancs\'{o}\inst{2} \and Orsolya V\'{a}s\'{a}rhelyi\inst{1} }
\authorrunning{Luis Natera et al.} 
%
\tocauthor{Luis Guillermo Natera Orozco, D\'{a}vid Deritei, Anna Vancs\'{o},
Orsolya V\'{a}s\'{a}rhelyi}
\institute{Central European University, Department of Network and Data Science, Budapest, N\'{a}dor utca 9. 1055, Hungary,\\
\email{Natera\_Luis@phd.ceu.edu},\\ WWW home page:
\texttt{https://networkdatascience.ceu.edu/}
\and
Corvinus University of Budapest, Budapest, Hungary}

\maketitle              

\begin{abstract}
Life quality in cities is deeply related to the mobility options, and how easily one can access different services and attractions. The pedestrian infrastructure network provides the backbone for social life in cities. While there are many approaches to quantify life quality, most do not take specifically into account the walkability of the city, and rather offer a city-wide measure. Here we develop a data-driven, network-based method to quantify the liveability of a city. We introduce a life quality index (LQI) based on pedestrian accessibility to amenities and services, safety and environmental variables. Our computational approach outlines novel ways to measure life quality in a more granular scale, that can become valuable for urban planners, city officials and stakeholders. We apply data-driven methods to Budapest, but as having an emphasis on the online and easily available quantitative data, the methods can be generalized and applied to any city.\makeatletter{\renewcommand*{\@makefnmark}{}
\footnotetext{{\small \papercitation}}\makeatother} 

\end{abstract}

\section{Walkability and liveable cities}

During the 20th century, most cities have evolved to accommodate a car-centric vision \cite{Jacobs1961}, allocating a privileged amount of urban space to motorized traffic \cite{Gossling2016,Szell2018}. From a liveability perspective, this situation is suboptimal because the automobile infrastructure dominates and defines the walkable area, increasing car traffic, air pollution and deteriorating walkable conditions.

The concept of walkability is an important factor to consider in connection with liveability. Liveability refers to an environment from an individual perspective \cite{Heylen2006} which includes "a vibrant, attractive and secure environment for people to live, work and play and encompasses good governance, a competitive economy, high quality of living and environment sustainability” \cite{Shamsuddin2012}. Thus in a liveable city, there must be an emphasis not only on sustainable transportation and built environment to reduce the harm on nature \cite{Campbell1996,Jabareen2013} but also encouraging citizens to walk for supporting their physical and mental well-being \cite{Frank2006}. However, improving walkability is more complex than we would think. Walking should be an available, safe and well-connected mode of transportation, but as Speck put it well, it should be interesting and comfortable as well, to have a feeling of the streets as ’outdoor living rooms’ \cite{Speck2012}.

The pedestrian infrastructure that sustains walkability in a city can be described as a network \cite{Porta2006}. This approach has been useful to identify street patterns \cite{Barthelemy2008,Louf2014b} and its evolution \cite{Strano2012,Barthelemy2013}, measure the morphology of cities \cite{Boeing2018}, and how the streets connectivity impacts on pedestrian volume \cite{Hajrasouliha2015}.

The various approaches to create a walkability index or so-called walk score consider mainly the following components: safety and security \cite{Quercia2015,Silva2018}; convenience, attractiveness and public policy \cite{Krambeck2006,Speck2012}, connectedness \cite{Southworth2005}, but also reckon with the land use mix and residential density of the certain area\cite{Carr2010}. Another approximation rather accents the importance of its effect on air pollution, health problems, travel costs and even on the sense of community\cite{Stephen2004}. Thus measuring walkability not only captures the propensity to walk in a city but also includes the components a liveable city must have and support, under the umbrella of sustainability.

There are good examples of how sustainable city development initiatives tackle growing inequalities with data-driven approaches. Long Island used city data to analyze which amenities are needed to increase the quality of life in a newly built environment \cite{LongIsland}, other cities are investing in smart technologies to develop public transport, connecting spatially discriminated areas \cite{NYC,Amsterdam}.

Since the number of components which should be taken into consideration in creating a walkability index is high, the types of data are also mixed and thus difficult to integrate. While the information on connectedness, security, residential density, etc. is quantitative and in general easily available, gaining opinion about attractiveness, convenience, or even about the feeling of security is more complicated. Here we propose to use a data-driven approach as a proxy to quantify life quality, making it reproducible and easily expanded to include different data sources. We apply our methods to Budapest, but as having an emphasis on the online and easily available quantitative data, the methods can be generalized and applied to any city.

\section{Data}
We work with three different data sources: networks, points of interest and city attributes. The pedestrian network and points of interest were acquired using OSMnx \cite{Boeing2017a}, a python library to download and construct networks from OpenStreetMap (OSM). The data contained in OSM is of high quality \cite{Haklay2010b,Girres2010} in terms of correspondence with municipal open data \cite{Ferster2019} and completeness: More than $80\%$ of the world is covered by OSM \cite{Barbosa-Filho2017}.

The majority of points of interest were downloaded from OpenStreetMap, from different classification keys (amenity, tourism, shop, office, leisure) using OSMnx \cite{Boeing2017a}. We filtered the points of interest using the districts' demarcation \cite{hu_dist}, to get only the data within Budapest boundaries, having, as a result, more than $39,000$ data points. We complement the data sets with secondary data sources as specialized directories of doctors and childcare facilities (see appendix).

We categorize the points of interest in six main categories: I) Family friendliness (Access to education and daycare, and family support services), II) Access to health care and sport facilities, III) Art and culture (e.g.: museums, exhibitions), IV) Nightlife (e.g.: bars, restaurants), V) Environment (air quality and access to green areas), and VI) Public Safety. The points of interest and secondary data sources are available at \url{https://github.com/nateraluis/Budapest_LQI}

The district-level data (population and crimes) were obtained from the Hungarian Police's public database, calculated based on the number of crimes committed in public places 100 thousand per capita in 2018 \cite{hun_police}. Population data is coming from the 2016 micro-census conducted by the Hungarian Statistical Bureau \cite{hun_pop}. We took into account the air pollution, this data set coming from National Air Pollution Measurement Network \cite{hun_env}, containing the geolocation of the air quality stations and different measures (annual median concentration of carbon monoxide, nitrogen dioxide, and PM10 dust).

Accuracy of the Life Quality Index (LQI) model highly depends on how comprehensive the distribution of listed services. We use OSM as our key data source, but to achieve a more comprehensive and country-specific database we collect publicly available data from various Hungarian websites for each category (See Appendix A for databases and sources)

The network contains all the sidewalks and pedestrian designated infrastructure, it is conceptualized as undirected, nonplanar and primal network \cite{Porta2006}. The pedestrian network is described as a weighted graph, with its adjacency matrix $W=w_{ij}$ where the weight $w_{ij}$ contains the length between $i$ and $j$ if connected, and $0$ otherwise.

We assigned properties to the nodes of the network, matching the nodes with their corresponding districts, then assign nodes as attributes based on the district level data (population and crimes, see section \ref{safety}). For the pollution data, we calculated the corresponding Voronoi cells, for the air quality stations, and matched the nodes with them, we divided the pollution by the number of nodes in each corresponding cell and assigned the value to the nodes (See section \ref{environment}). For the edges, we encoded their length $\ell_{ij}$ along with the traversal time $Tt_{ij}$ between nodes $i$ and $j$ calculated as $Tt_{ij}=\frac{\ell_{ij}}{ps}$ where $ps$ is the pedestrian speed as a constant rate of $5km/h$.

\begin{figure*}[htbp]
	\centering
	\includegraphics[width=0.8\textwidth]{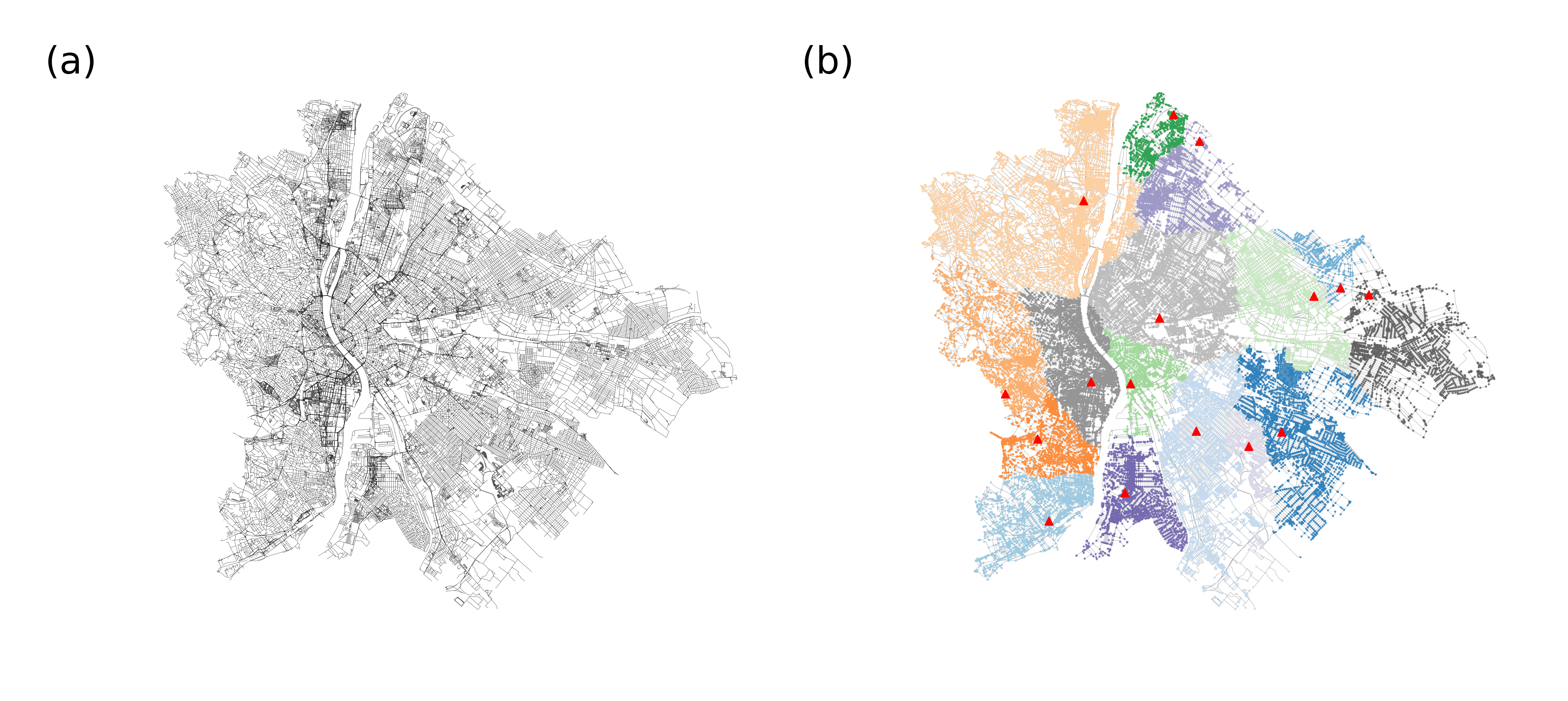}
	\caption{\textbf{(a)} Network representing Budapest pedestrian structure. The network was built following a primal approach, where the edges are sidewalks and pedestrian infrastructure, and nodes are intersections. \textbf{(b)} The graph-Voronoi tessellation of the Budapest network, generated using a subset of 15 parks as seeds. The color of the nodes represents the cell they belong to and the highlighted red dots are the seeds of each cell. The distance measure between two points is defined as the weighted shortest path on the graph, the weights being the average time required to cross a given edge.}
	\label{fig:BPnetwork}
\end{figure*}

\section{Quantifying Life Quality}
The life quality of a person is largely subjective and hard to quantify. However, it is both intuitive and has been scientifically shown that the environment and personal well-being strongly correlate \cite{rosow_1961}. Thus using environmental factors as proxies, life quality and livability becomes quantifiable \cite{kahneman_2006}.

The main environmental factors we consider in our model are: the availability of services and amenities, the quality of the infrastructure, environmental factors and safety. The goal of our model is to quantitatively characterize the immediate environment of residents in the space of factors that affect life quality.

The fundamental framework of our model and our calculations is the network representation of Budapest’s pedestrian infrastructure. The nodes of the network represent intersections, while links are sidewalks and pedestrian infrastructure. The output of our model is an index, that characterizes every node of the Budapest network, giving a high-resolution quality-landscape of the city. The index is ultimately a number aggregated from multiple sub-categories, and its main value is highlighting inequalities and relative deficiencies within the city.

The final value of the index is a weighted sum, characterizing every node (intersection) in the network:
\begin{equation} \label{final_Q}
    Q_i =w^{services}\Tilde{Q}_i^{services} + w^{safety} \Tilde{Q}_i^{safety} + w^{environment}\Tilde{Q}_i^{environment}
\end{equation}
In the equation $i$ represents an individual node in the network. The \say{tilde} above the $Q$ terms means that the values of the different category indices are normalized within the category. The weights $w$ assigned to every term are arbitrary and are highly context-dependent. We include the weights used for producing the results of this paper in Appendix B.  All terms of the equation are discussed in the following sections.

\subsection{The services index: $Q^{services}$}
The number quantifying each node in terms of how well it is connected with amenities and services is a weighted sum of sub-categories as well.
\begin{equation}\label{Q_services}
    Q_i^{services} =\sum_c w^c Q_i^c
\end{equation}

where, $c$ denotes categories (family, culture, health, sport, and nightlife), and $w^c$ the importance (weight, see Appendix B) of category $c$.
Some categories, like family, have further subcategories. Even though we have also had data and made a separate analysis on tourism, its effects on life quality of the residents are ambiguous, so we decided to omit it from the index.\\

What sets categories apart is that they incorporate different sets of amenities, with a few overlaps. The details of the categorization of amenities are included in the Appendix. \\

For every service/amenity class we have a given set of points of interest (POI) along with where the amenities of that class are available, with exact geo-location. We assign every POI of a given amenity class (e.g supermarket, pharmacy, school, etc.) to the nearest node on the infrastructure network. Each set of POIs organically generates a spatial partitioning of the city with one partition per POI. The partition of a POI is the set of all the nodes from which that particular POI can be reached faster than any other POI of the same class.

Mathematically these partitions are called graph-Voronoi cells \cite{erwing_2000,Deritei2014}, where every node of a cell is assigned to its closest seed (POI). Distance, in this case, is not euclidean or geometric distance, but the distance on the network, where we use the weighted shortest path between two nodes as the distance measure. The weight of links is a temporal parameter encoding the average time required to cross the represented street from one end to the other, thus the weight is a simple product of average speed and length of the street. This is in principle very similar to the way navigation systems find routes between points. For an example of a graph-Voronoi partitioning see Figure 1 (b).\\
To assess how well connected a node is to amenities we consider the following factors:
\begin{itemize}
    \item How important is an amenity - weight ($w_a$)
    \item How long does it take to reach the amenity - time to reach ($t_ia$)
    \item Relatively how many nodes (or people) does the amenity share with - exclusivity ($P_a$)
\end{itemize}

From the three factors, the latter two are calculated using the city infrastructure network. The index for an amenity class, from the perspective of node $i$, is proportional with its importance (weight) and it is inversely proportional with the time to reach the closest POI from $i$ and with the degree of exclusivity.
\begin{equation}\label{q_i}
    q_i^a=\frac{w_{a}}{(P_a+1)(t_{ia}+1)}
\end{equation}

There can be certain singular cases when a Voronoi cell is empty ($P_a=0$, i.e. no residents in the area) or the node i in question is right at the POI ($t_{ia}=0$). To avoid anomalies in the index we added 1 to both parameters.

The index of one category is proportional to the sum of its amenity-indices (calculated in (\ref{q_i})). To treat this number on the right scale (in practice we can get very large and very small numbers) we take the natural logarithm of the sum across amenities.
\begin{equation}
    Q_i^c=log(\sum_a q_i^a):
\end{equation}

As we have mentioned earlier the final services index is the weighted sum of the indices of the sub-categories.
$$Q_i^{services} =\sum_c w^cQ_i^c $$
Finally we normalize the values of $Q^{services}$ so its values are comparable to the other values of the final $Q$ equation (\ref{final_Q}):
\begin{equation}
    \Tilde{Q}_i^{services} =\frac{Q_i^{services}+|min(Q^{services})|}{max(Q^{services})+|min(Q^{services})|}
\end{equation}

\subsection{Safety index: $Q^{safety}$} \label{safety}

The safety index is calculated across districts based on the number of crimes committed per one hundred thousand residents. Since the highest resolution data available to us was on the district level, every node $i$ in the same district will have the same safety index value. The crime index:

$$Q_i^{crime}=\frac{N^{district}_{crime}}{n_i^{district}}$$
Where $N^{district}_{crime}$ is the number of crimes committed in a district in a year, and $n_i^{district}$ is the number of nodes in the district. The safety index is one minus the normalized crime index.
\begin{equation}
    \Tilde{Q}_i^{safety}=1-\frac{Q_i^{crime}}{max(Q^{crime})}
\end{equation}

\subsection{Environmental index $Q^{environment}$} \label{environment}
The environmental index is made up of two components: air pollution ratio and ratio of natural areas.

\subsubsection{Air pollution ratio}
We use the data provided by Budapest’s air pollution measuring stations for the year 2018. For this study, we used the yearly median value of three polluters: carbon monoxide, nitrogen dioxide, and PM10 dust-pollution.
As an approximation, we project the geometric Voronoi cells of the measuring stations onto the city map and each node will receive the pollution metrics of the geometrically closest station. We divide these values with the yearly upper health limit for the given polluter to assess to what degree do these values approximate the health limit. Thus the air pollution index of one node is formalized as follows:
$$C_i=\frac{c_i^{CO}}{c_{limit}^{CO}}+\frac{c_i^{NO_2}}{c_{limit}^{NO_2}}+\frac{c_i^{Pm10}}{c_{limit}^{Pm10}}$$

Where $c^{CO}_{limit}=3000 g/m^3$, $c^{NO_2}_{limit}=40 g/m^3$ , $c^{Pm10}_{limit}=40 g/m3$ are the yearly upper limits based on \cite{levegominoseg}.

\subsubsection{Ratio of natural areas}
For this index, we have data on the neighborhood level, which is a more granular level of administrative partitioning the city than the districts are. In this case, we project the same index onto every node in the same neighborhood. We consider as natural areas forests, parks and water surfaces (ponds, rivers, etc). \\
The index:
$$T_i=\frac{R_{water}^{nh(i)}+ R_{forest}^{nh(i)}+R_{park}^{nh(i)}}{max(T)}$$
Where $R_x^{nh(i)}$ is the relative surface area of natural area $x$ within the neighborhood that $i$ belongs to ($nh(i)$). In other words, the surface area of a natural area is divided by the number of nodes in the neighborhood and the surface area of the neighborhood. Thus
$R_x^{nh(i)}=\frac{T(x)}{T(nh(i))n_{nh}}$, where $T(x)$ is the surface area of $x$ natural area, $T(nh)$ is the surface area of $nh$ neighborhood and $n_{nh}$ is the number of nodes in neighborhood $nh$.
The final environmental index:
$$Q_i^{environment}=\frac{1+T_i}{1+C_i}$$,
That after a normalization is:
$$\Tilde{Q}_i^{environment}=\frac{Q_i^{environment}}{max(Q^{environment})}$$

\section{Results}
We quantify life quality in terms of each category (family support, education healthcare, sport, culture, nightlife, environment), and an overall measurement which contains all 6 categories and crime rate normalized by the population for the city of Budapest. Our method allows us to measure life quality for each intersection of the city, which helps to capture within neighborhood inequalities too. Analysis on the category level is beneficial for targeted policy interventions for better service allocation.

\begin{figure*}[htbp]
	\centering
	\includegraphics[width=0.8\textwidth]{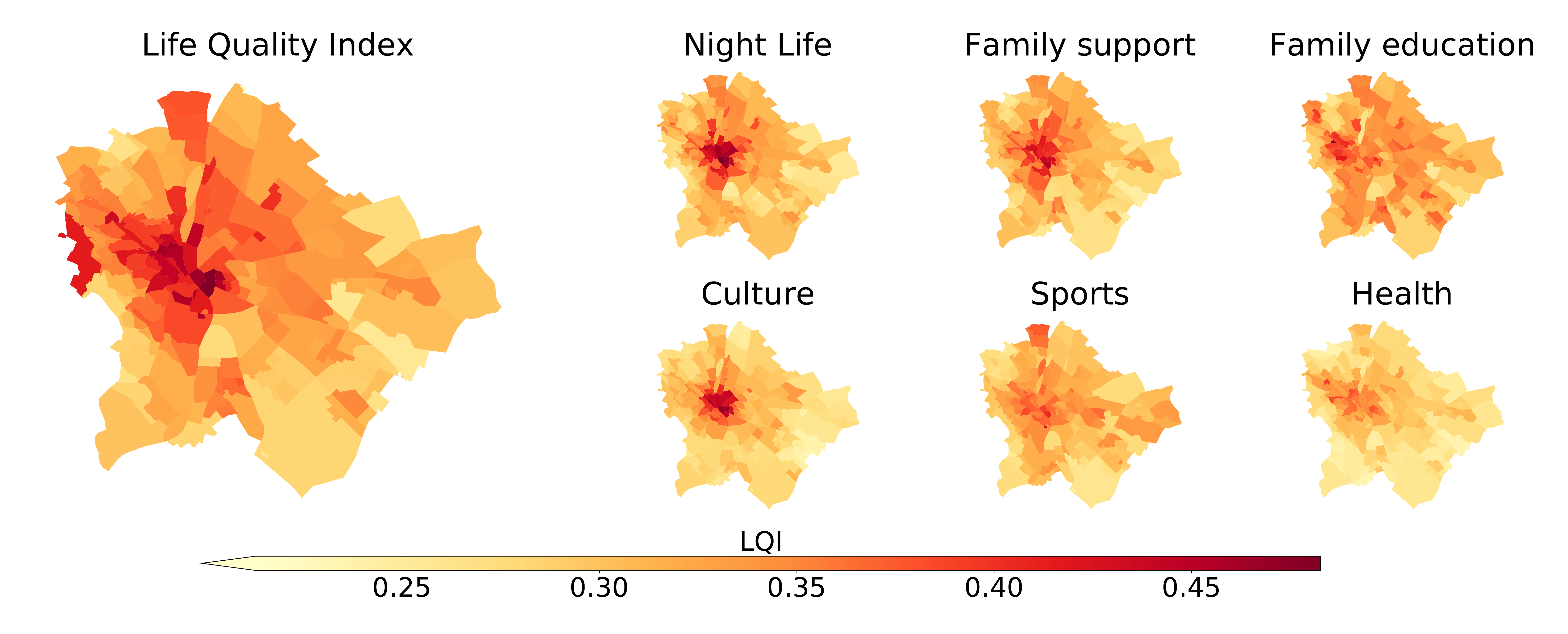}
	\caption{Budapest neighborhoods, average life quality by categories and aggregated life quality index.}
	\label{fig:BPlqi}
\end{figure*}

Figure \ref{fig:BPlqi} shows our overall life quality index (LQI) and by categories. Heatmaps reveal important features of Budapest. Similarly to most European cities life quality is much better in the inner districts \cite{Hohenberg1986,Brueckner1999}, especially in the case of Night Life and Culture.

Budapest is divided by the Danube river into two main parts: Buda and Pest. The river does not only serve as a geographical border but due to historical reasons, it also divides the citizens by social status. Hilly Buda, on the West side of the river, used to be the capital of the country, with the residence of the former Hungarian king. On the other side, the mainly flat Pest used to be the agricultural supporter of the aristocrats in Buda \cite{Gyani}. Even though the city has changed dramatically since the Monarchy, the division of Buda and Pest persists, and our life quality index captures it well. However certain services are legally guaranteed to be evenly distributed in the city, such as education and healthcare, for precise modeling one should take into account private care too, which highlights inequalities. So, the traditional division of Buda and Pest is even visible in categories where there should not be that much of a difference (Education, Family Support, Healthcare).

Results also highlight that category LQI-s are highly correlated, less liveable neighborhoods are constant regardless of the amenity category, and well-performing neighborhoods do not change either. It is caused by two main factors: the lack of amenities and the relatively high walking distances in the suburbs.

The compact city concept focuses on building more sustainable and livable cities while designing practical neighborhoods where citizens can maintain everyday life without a car \cite{Dittmar2012}. Since, the walkability of a neighborhood highly correlates with its liveability \cite{Rogers2011} and the suburbs in Budapest do not show any compact city design features, both long distances and the lack of amenities effects suburban habitats lives negatively.

\subsection{Evaluation}

Multiple methods have been developed to evaluate the accuracy of quality of life metrics: Scholars used expert validation with geographic visualization \cite{Rinner2007,Gavrilidis}, correlations with socioeconomic characteristics \cite{Talen} and surveying citizens’ perceptions of the conditions of life \cite{Santos}.

Our evaluation is based on the micro-economical {\it hedonic approach} of estimating the values of public goods. In a capitalist market, real estate prices reflect the recognition of a neighborhood's characteristics: Prices are formed based on demand, more desirable places are more expensive, due to the underlying assumption of providing a higher quality of life. \cite{Brueckner1999} Estimating neighborhoods life-quality with real estate prices has a long tradition in urban literature \cite{Roback,Hoehn,Lora}, therefore we adopt this method to evaluate our model.

\begin{figure*}[htbp]
	\centering
	\includegraphics[width=0.8\textwidth]{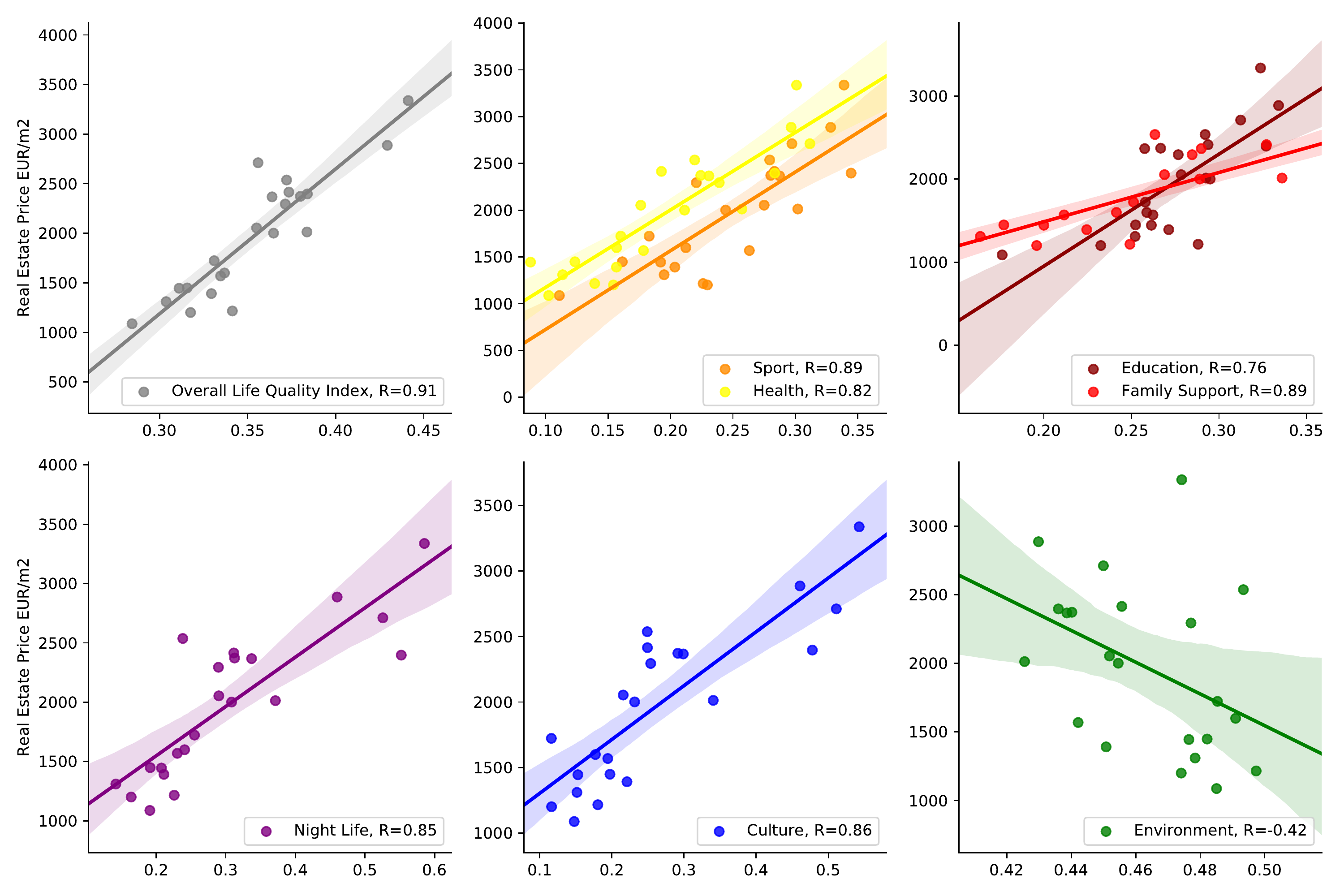}
	\caption{Districts of Budapest Life Quality Index (LQI) and its components correlated with real estate prices ($m^{2}/EUR$)}
	\label{fig:LQI_ev}
\end{figure*}

We collected the average $m^{2}/EUR$ price for all 23 districts of Budapest in January 2019 \cite{real_price} and correlated each LQI category averaged by district with it. Figure \ref{fig:LQI_ev} shows that our overall LQI correlates the most (R=0.91) with the real-estate prices. Most of its components have a positive correlation with real-estate prices, except the environment which is calculated based on air pollution and green surface proximity. The life quality (LQI) in Budapest is much higher in densely populated downtown districts, which are lack of green surface and suffers from high air pollution due to heavy traffic.

\paragraph{Summary} Locals of Budapest, like in most European cities, traditionally values downtown areas. The relative closeness to CBD, good access to public transport, and vital city life kept it as a desirable area for living \cite{Cassiers2012}. However, in recent years, the city is facing new challenges: due to gentrification \cite{Garcia} and over-tourism (eg.:Airbnb) real estate prices are sky-rocketing in downtown areas. In contrast with the early 2000-s when (upper) middle-class moved to the suburbs, nowadays, lower-income families and young professionals are leaving the downtown behind in hope for more affordable living.

As our findings show, Budapest is quite centralized and the quality of life highly correlates with real estate prices, which possibly lead to even more inequalities in the future. This spatial discrimination with longer traveling time, less fulfilling environment, and potential segregation reduces the chances of upward mobility and the quality of life of individuals \cite{Gobillon}.

\section{Discussion}
We have proposed a methodology to quantify life quality as a function of walkability on urban networks. We have used open data to capture inequalities between neighborhoods and districts in the city. We have shown that the real estate market reflects the life quality that our methods found.

A data-driven approach for quantifying life quality at such a granular level like our proposed method can help decision-makers to tackle social and environmental challenges better. Designing compact, liveable neighborhoods, considering also the upcoming environmental crisis is the number one priority of many cities worldwide.

The use of open data sources and algorithmic approaches adds up towards a systematic framework for understanding urban liveability. Our current approach is not the last word in this development since it does not yet account for multiple other variables, such as the quality of services and infrastructure, and other qualitative variables. To capture the more specific indicator of liveability in different cities it would be necessary to work with more granular and city-dependant data.

We anticipate a future stream of research focused on the use of worldwide open data sets to quantify urban liveability, including longitudinal studies in multiple cities, along with algorithmic modeling, simulations, and machine learning approaches, to first quantify the liveability, propose changes and test them with the ground truth data.

\section*{Acknowledgments}
The authors wish to thank the experts of KKBK for consultations, and  Federico Battiston and Gerardo I\~{n}iguez for comments and discussions on the subject.\\

\section*{Appendix}
\appendix
\section{Secondary Data sources}
\begin{itemize}
    \item {Sport associations in Budapest}\cite{hun_sport}
    \item {Kindergartens, daycares, primary and secondary education}\cite{hun_edu}
    \item {Art and music schools}\cite{hun_art}
    \item {Child health services} \cite{hun_child_health}
    \item {Social welfare system (eg.: elderly care)} \cite{hun_social}
    \item{Culture centers} \cite{hun_cultcent}
    \item {Indoor playgrounds} \cite{hun_playin}
    \item{Healthcare (hospitals, private and public clinics, specialists)} \cite{hun_health}
    \item{Fitness and training facilities} \cite{hun_fitness}
    \item{Outdoor fitness facilities} \cite{hun_outfitness}
    \item{Thermal baths and spa} \cite{hun_thermal}
    \item{Playgrounds and parks} \cite{hun_parj}
\end{itemize}

\section{Weights used in the calculations}
The weights of the different $Q$ indices in the final aggregation as well as in sub-categories highly depends on the context and the nature of the problem. Here we present the values we used to generate the results of this study, that were agreed upon consulting with experts.
The weights of the sub-indices from equation (\ref{final_Q}) are of the following values:\\
$w^{services}=0.7$\\
$w^{safety}=0.1$\\
$w^{environment}=0.2$\\
The category weights used in equation (\ref{Q_services}), aggregating $Q^{services}$ are:\\
$w^{family}= 0.3$;\\
$w^{health}= 0.3$;\\
$w^{culture} = 0.15$;\\
$w^{sport} = 0.15$;\\
$w^{night life}=0.1$;\\


\end{document}